\documentclass[aps,
 pra,twocolumn,
 amsmath,amssymb
]{revtex4-1}
\usepackage{graphicx}
\usepackage{dcolumn}
\usepackage{bm}
\usepackage{amsmath,amssymb,graphicx,amsbsy,multirow,tabu,braket}
\usepackage{hyperref}
\graphicspath{{images/}}

\begin{document}

\title{Manipulating matter waves in an optical superlattice}

\author{Brendan Reid\textsuperscript{1}, Maria Moreno-Cardoner\textsuperscript{1}, Jacob Sherson\textsuperscript{2}, Gabriele De Chiara\textsuperscript{1}\\
 \textsuperscript{1}{\it Centre for Theoretical, Atomic, Molecular \& Optical Physics, Queen's University, Belfast BT7 1NN, Northern Ireland}\\
\textsuperscript{2}{\it Department of Physics and Astronomy, Ny Munkegade 120, Aarhus University, 8000 Aarhus C, Denmark.}
}
\begin{abstract}
We investigate the potential for controlling a non-interacting Bose-Einstein condensate loaded into a one-dimensional optical superlattice. Our control strategy combines Bloch oscillations, originating from accelerating the lattice, with time-dependent control of the superlattice parameters. We investigate two experimentally viable scenarios, very low and very high potential depths, in order to gain a better understanding of matter wave control available within the system. Multiple lattice parameters and a versatile energy band structure allow us to obtain a wide range of control over energy band populations. Finally, we consider several examples of quantum state preparation in the superlattice structure that may be difficult  to achieve in a regular lattice.
\end{abstract}

\maketitle


\section{\label{intro}Introduction}
The simulation of quantum processes using optical lattices has attracted intense research into their advantages and viability in recent years \cite{Blochnature, Cirac2012, Buluta2009}. While a universal quantum simulator as envisioned by Feynman \cite{feynman} may still be a number of years away, many important advancements have been made.  Having complete control over the structure of an optical lattice, coupled with a wide range of applications, has put these types of systems at the forefront of quantum simulation research. Applications have been found in  relativistic field theories for fermions \cite{Mazza2012}, exotic forms of magnetism \cite{Simon2011}, implementation of quantum logic gates \cite{qubitgate} and demonstration of phase transitions from a superfluid to a Mott insulator in ultracold atoms \cite{Greiner2002}. Engineering specific quantum states and exercising control is vital for quantum simulation and ultracold atoms loaded into optical lattices have shown promise in this regard \cite{gdc, adc}. The matter-wave nature of ultracold atoms means Bose-Einstein condensates (BEC) can be reliably controlled in an optical lattice \cite{controlref} and prepared in non-trivial states desirable for experimental consideration \cite{muller}. These systems have been used in spin-exchange interactions \cite{trotzky}, exact control of the atomic number distribution using the interaction blockade mechanism \cite{mariona} and a promising platform for investigating frustrated geometries \cite{kagome}. Similar to the work performed in \cite{park}, here we employ Bloch oscillations as a driving force for particles, highlighting the potential for arbitrary generation and coherent control of quantum states in a one-dimensional superlattice structure.

Bloch oscillations were originally formulated to describe electron motion in crystalline structures in the presence of an external electric field \cite{bloch1929}. Maintaining these crystals in a regime where quantum effects can occur is experimentally difficult and so periodic optical lattices have been employed as a substitute, with the experimental observation of Bloch oscillations confirmed in Refs. \cite{Dahan,Wilkinson}. Replacing crystals with optical structures does not change the nature of Bloch oscillations: any external force acting on a periodic system, such as gravity in a vertical lattice or inertial forces in accelerated lattices, will induce these oscillations \cite{morsch2,Anderson}. Optical waveguides are an alternative platform to observe Bloch oscillations and other coherent wave phenomena \cite{Pertsch,waveguides}. 

Control of the vibrational states of a BEC in a conventional (or simple) optical lattice has been achieved using Landau-Zener tunnelling \cite{MorschNaturePhysics, referee3, zenesini2009} and more sophisticated optimal control techniques to generate trapping potentials in atom chips \cite{vonfrank}. Optimal preparation of the internal state of ultracold atoms trapped in optical lattices and atom-chip devices has also been achieved \cite{Zibold, Lovecchio}.  The interesting combination of spin-dependent forces and Bloch oscillations allow one to perform quantum simulation of relativistic effects \cite{Witthaut}. 
\begin{figure}[t]
 \centering
\includegraphics[width=\columnwidth]{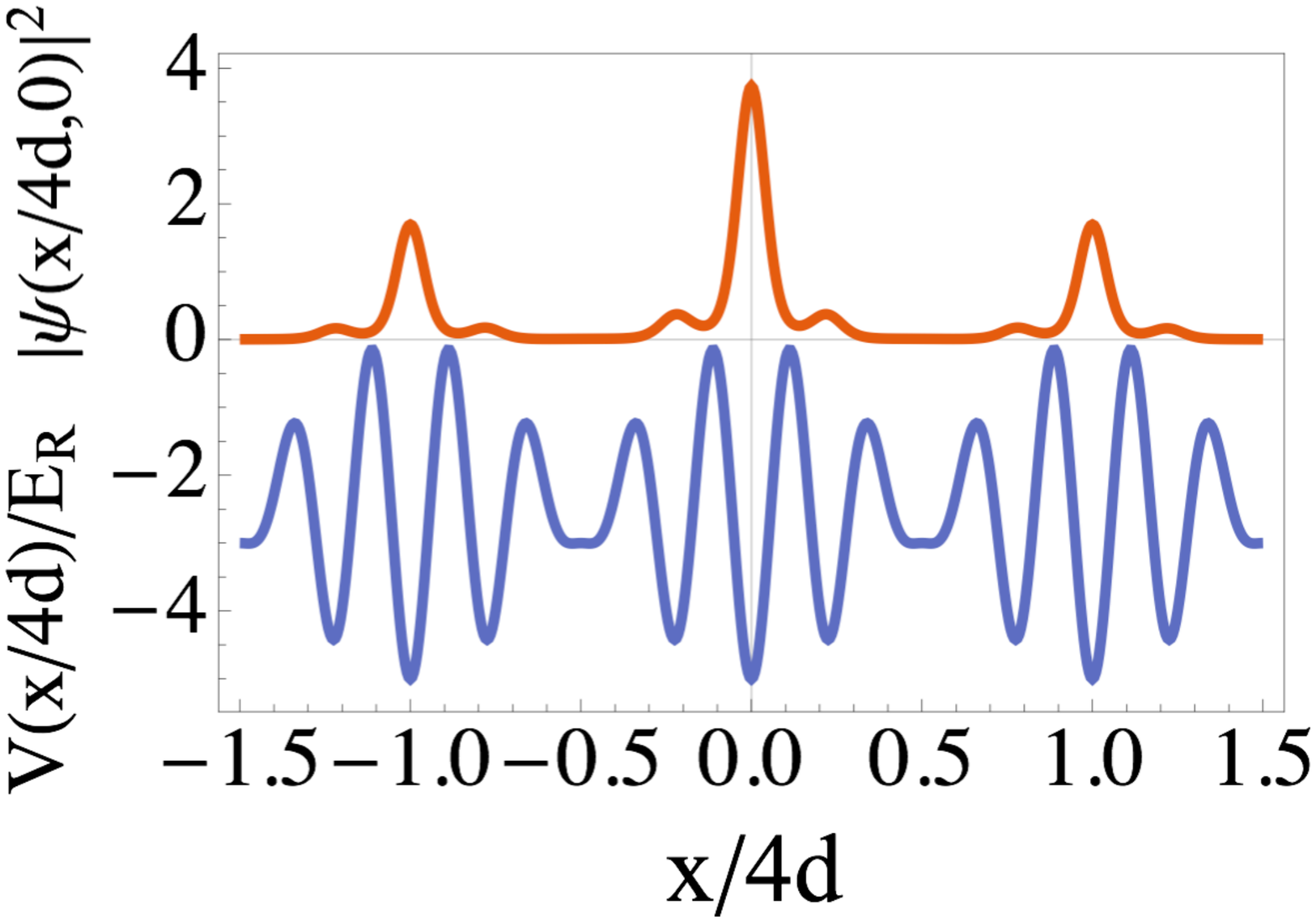}
 \caption{\label{iwf}{\it Colour online.} The top curve shows the modulus square of the initial wave function \ref{iwf} localised into the lowest energy band with an initial quasimomentum distribution with standard deviation $\sigma=2k_r/5$. The bottom curves is the potential \ref{potential} plotted on the same scale. Here the superlattice parameters are $A_1=3E_R$, $A_2=2E_R$ and $\phi=0$. }
\end{figure}
Combining Bloch oscillations with a non-standard lattice structure, a superlattice, allows us to investigate control processes that may not be possible in a simple lattice \cite{bsuperlattice, gsuperlattice}. 

\begin{figure*}
       \includegraphics[scale=.3]{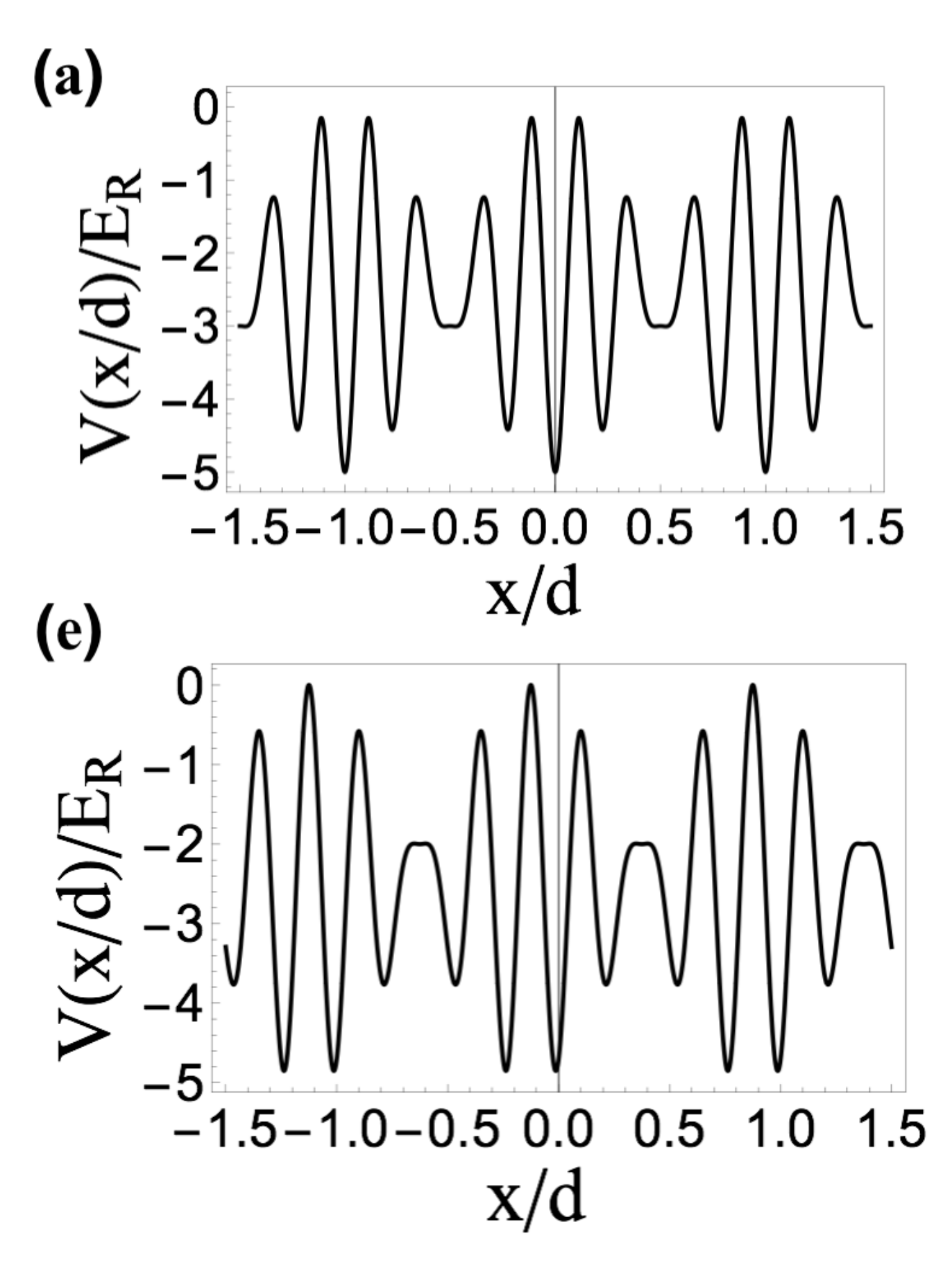}
        \includegraphics[scale=.31]{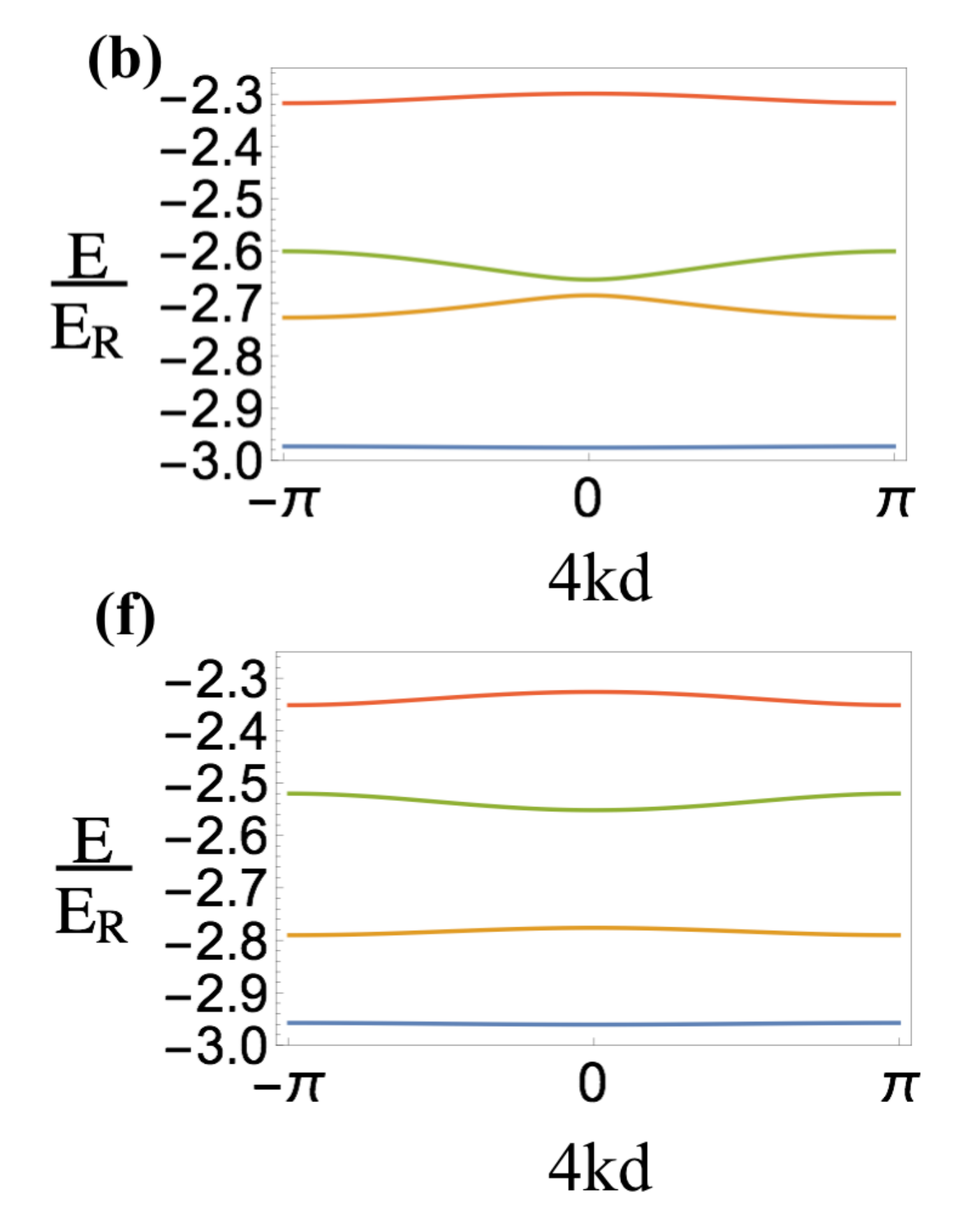}
       \includegraphics[scale=.32]{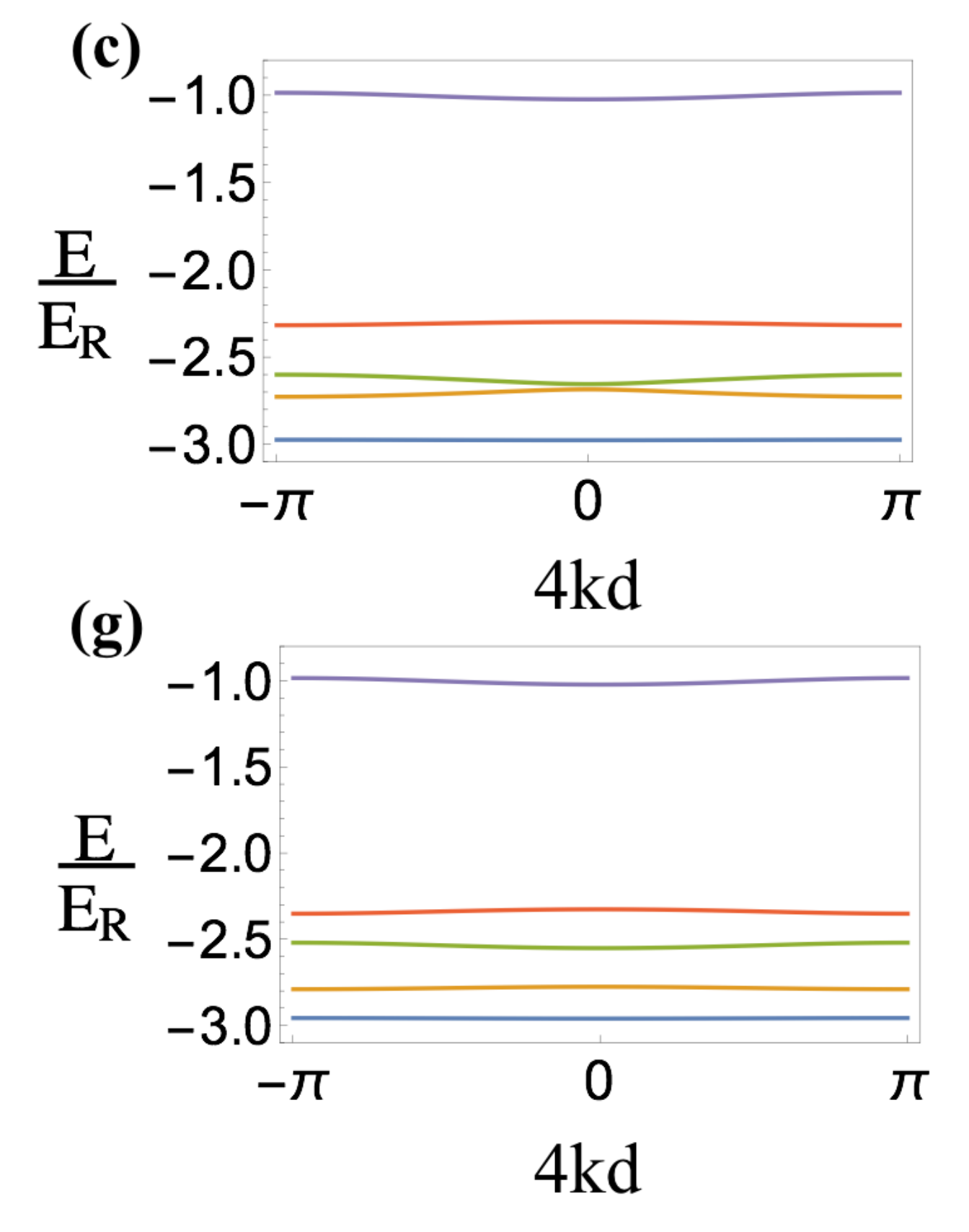}
\includegraphics[scale=.31]{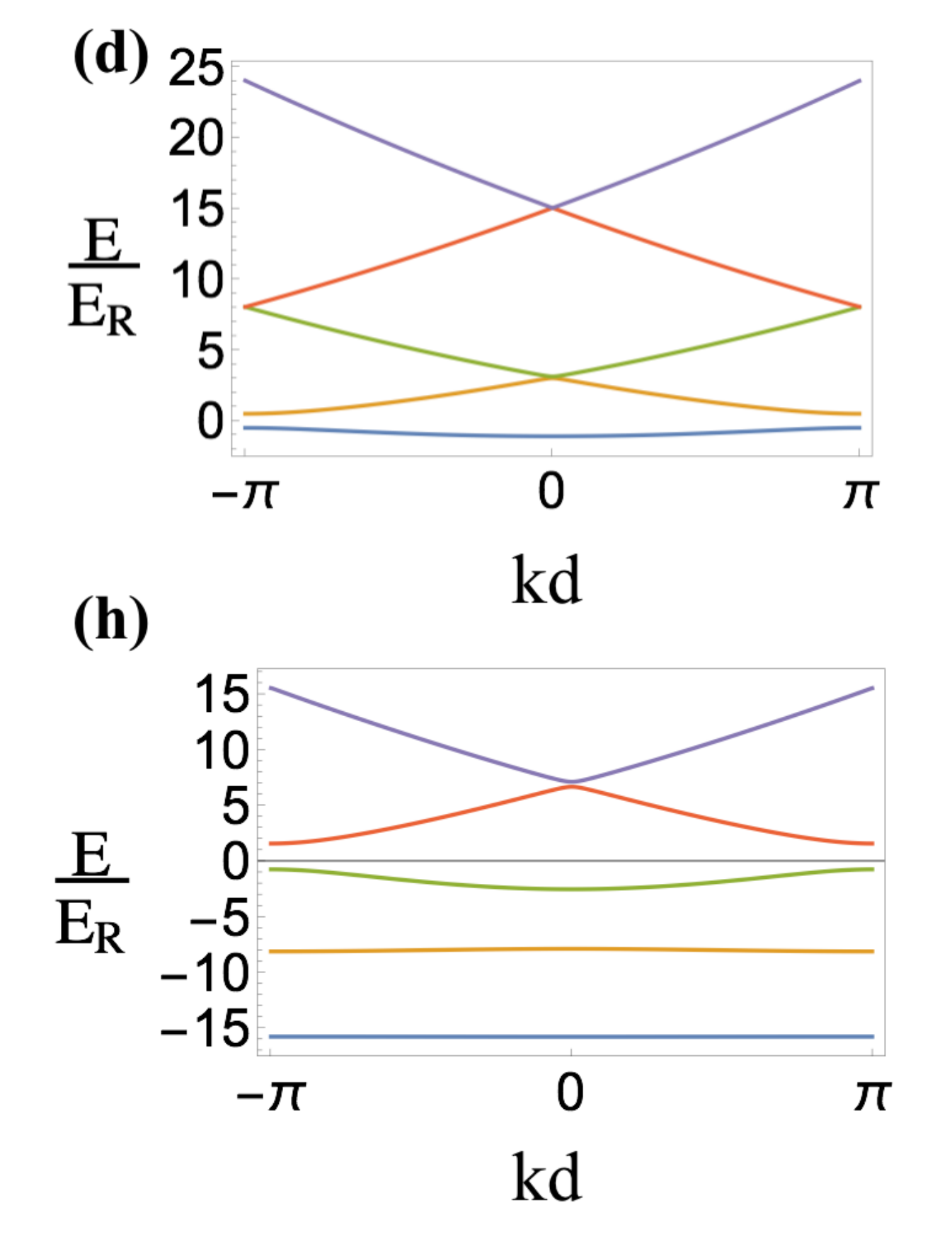}
\caption{\label{bspotcomp}{\it Colour online.} 
(a) Superlattice  plotted in real space. (b) Lowest four and (c) five energy bands plotted in momentum space. Here $A_1=3E_R$, $A_2=2E_R$ and $\phi=0$. (e)-(g) have the same potential depths but $\phi=\pi/8$. (d) and (h) display the band structure for the simple lattice when $V_0=2E_R$ and $V_0=20E_R$ respectively. }
\end{figure*}
In this paper we consider a period-4 superlattice obtained from combining two optical fields with close-lying wavelengths \cite{benpaper}. Such a lattice exhibits an interesting energy band structure: the gaps between bands do not decrease monotonically with band index. The transport of atoms, caused by Bloch oscillations, through such an interesting band structure provides an experimentally viable landscape for complex quantum state preparation. In contrast to previous approaches for the control of atomic wave packets, we combine Bloch oscillations with time-dependent control of the superlattice, achieved through step-wise changes of the lattice parameters, with the aim of manipulating a noninteracting BEC and creating non-trivial superpositions of momentum states in different bands.

The paper is organized as follows. In Sec. \ref{methods} we present the theoretical methods behind our numerical calculations and simulations. Section \ref{superbo} goes into detail on the effect Bloch oscillations have on wave packets inside a periodic lattice structure. Section \ref{manipulate} details the control processes we employ to generate and manipulate quantum states of matter within the superlattice structure. In Sec. \ref{conc} we summarise.

\section{\label{methods}Methods}

Bloch's theorem states that translational invariance in a periodic potential causes particle eigenstates to have well defined quasimomentum (or crystal momentum). Thus, for particles in an optical lattice, a description in quasiquasimomentummomentum space is often more convenient than in real space. Applying an external force to a particle in an optical lattice will increase its quasimomentum linearly in time and proportionally to the magnitude of the force.
A condensate with zero initial quasimomentum will therefore begin to travel through the lattice. Due to the periodicity and symmetry of the potential structure, the quasimomentum dynamics can be redefined in the first Brillouin zone. This movement in quasimomentum space manifests as a periodic and symmetric evolution in position space as explained in \cite{dynam}. The amplitude of these oscillations is inversely proportional to the force, with the period given by $\tau_B=2\hbar \pi/d_{lat}F$ where $d_{lat}$ is the spatial period of the lattice considered. As $\tau_B$, known as the Bloch period, becomes smaller, inter-band transitions can occur at any avoided crossings present in the band structure. We can use these crossings to split a BEC into a superposition of states \cite{breid1}.

While the solutions for the Schr\"{o}dinger equation with a periodic potential are well known, the addition of the force means that Bloch functions are no longer eigenstates of the Hamiltonian. In order to solve the one-dimensional Schr\"{o}dinger equation for the Hamiltonian 
\begin{equation}\label{h}
H=-\frac{\hbar^2}{2m}\partial_x^2 +V(x)-Fx
\end{equation}
where $V(x)$ is a periodic function, we use an ansatz of the real space wave function \cite{kolovsky},
\begin{equation}\label{wf}
\psi(x,t)=\sum_{\alpha} c_{\alpha}(k)e^{-\frac{i}{\hbar}\int_0^tdt^\prime E_\alpha(k^\prime)}\Phi_{\alpha,k}(x).
\end{equation}
The index $\alpha$ is a band index, $E_\alpha(k)$ denotes the energy value for the band $\alpha$ at quasimomentum $k$, $\Phi$ are the Bloch functions of the optical lattice and $k^\prime$ is the time-dependent quasimomentum defined 
\begin{equation}\label{timequasirel}
k^\prime= k_0+\frac{F t^\prime}{\hbar}
\end{equation} for each $k_0=k(t=0)$ across the Brillouin zone. 

The coefficients $c_\alpha(k)$ are found by solving the differential equation
\begin{equation}\label{cterm}
\dot{c}_{\alpha}=-\frac{i}{\hbar}F\sum_\beta\int dx X_{\alpha,\beta}(x) \hspace{5pt} e^{-\frac{i}{\hbar}\int^t_0 dt^\prime \hspace{3pt}\Delta_{\alpha,\beta}}c_{\beta},
\end{equation}
where $X_{\alpha,\beta}(x)=\Phi^*_{\alpha,k}(x)\partial_k \Phi_{\beta,k}(x)$ and $\Delta_{\alpha,\beta}=E_\alpha(k^\prime)-E_\beta(k^\prime)$.  The initial state we consider, $\psi(x,0)$, determines the initial conditions $c_\alpha(k_0)$. The solutions to Eq.(\ref{cterm}), $\mathbf{c}(t)=\{c_\alpha(t)\}_{\alpha=1,...}$, provide information on how an atomic cloud behaves in a periodic band structure, including how it interacts with avoided crossings.
The absolute square value of the solutions can be interpreted as the `population probability' of each energy band. Periodic potentials in the presence of an external static force can also be described in terms of Wannier-Stark resonances \cite{glucknew} providing an alternative framework to study wave packet propagation \cite{hartmannnew}.  

The periodic lattice system that we will consider is a superlattice ---  an incoherent sum of two lattices realised with different light polarisations. The resulting potential is given by
\begin{equation}\label{potential}
V(x)=-A_1 \cos^2(k_1 x)-A_2 \cos^2(k_2 x+\phi).
\end{equation}
 We consider a specific case of Eq.(\ref{potential}), where we have defined the wave vector $k_2=5 k_1/4$. This choice of wave vector for the secondary lattice creates a superlattice structure with a periodicity four times that of a simple lattice with wave vector $k_1=\pi/d$, $d=\lambda/2$ where $\lambda$ is the wavelength. The recoil quasimomentum (or half-width of the first Brillouin zone) of the superlattice is $k_r=\pi/d_{super}$, where $d_{super}=4d$. The potential depths $A_1$ and $A_2$ are measured in the usual units of recoil energy for a single lattice $E_R=\hbar^2\pi^2/2md^2$. We are using a matter wave to simulate a non-interacting BEC of \textsuperscript{87}Rb atoms trapped inside this lattice, $m \equiv m_{Rb}$. If the BEC is prepared initially localised to the lowest energy band, with a Gaussian quasimomentum distribution centred on $k=0$ and standard deviation $\sigma=2k_r/5$, Fig.\ref{iwf} shows the modulus square of the initial wave-function $\psi(x,0)$ with the potential structure $V(x)$ plotted below on the same axis, displaying how the peaks of the atomic density distribution correspond directly to the wells of the superlattice.

The relationship between the wave vectors of the two lattices produces the non-standard structure of the superlattice. The superlattice parameters $A_1$ and $A_2$ can be used to control how deep the potential is and the relative phase $\phi$ affects the layout of the unit cell, changing the deepest minimum in the potential from singly to doubly degenerate ($\phi=0\rightarrow \pi/8$). This modification of the lattice in real space has a corresponding effect in momentum space. With no relative phase present the second and third energy bands are closer in energy, approaching degeneracy for very high potential depths (see Sec. \ref{highpot}). Increasing the relative phase up to $\pi/8$ increases the energy difference between the second and third energy bands, moving each band closer in energy to the first and fourth bands respectively. For a more complete picture of how the relative phase affects the distances between bands in the superlattice, Fig.1 in the Supplementary Material \cite{supp} gives more details.

 Figure \ref{bspotcomp} shows the potential structure and the corresponding band structure of the lowest four and five energy bands. In the regime of $A_1>A_2$ the energy difference between the fourth and fifth energy bands is much larger than the interband gaps between the first four bands, while in the regime $A_1<A_2$ the fifth band decreases in energy, moving closer to the lowest four, and a large gap opens up with respect to higher energy bands. In each case the superlattice has four and five wells respectively in its unit cell, whilst its periodicity remains un-changed. 
Figures (d) and (h) in Fig.\ref{bspotcomp} show the band structure of a simple lattice of the form $V(x)=-V_0\cos^2(\pi x/d)$ for comparison. In stark contrast to the superlattice, the band gaps for the simple lattice decrease monotonically with the band index and a quasi-isolated set of bands within which one could perform state engineering is not available. This highlights the usefulness of the superlattice.

%
\begin{figure}[t]
 \centering
 \includegraphics[scale=.26]{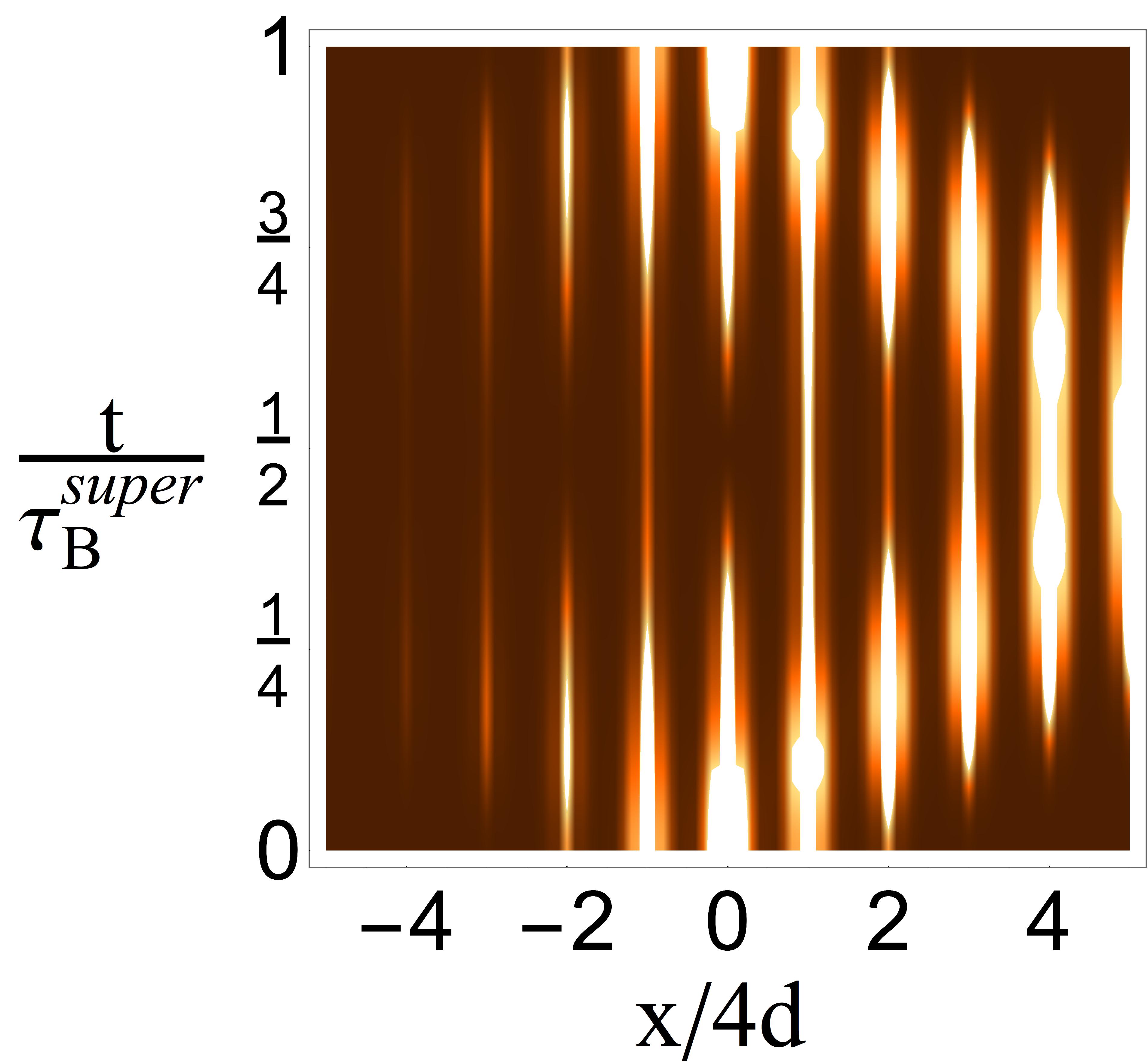}
  \includegraphics[scale=.26]{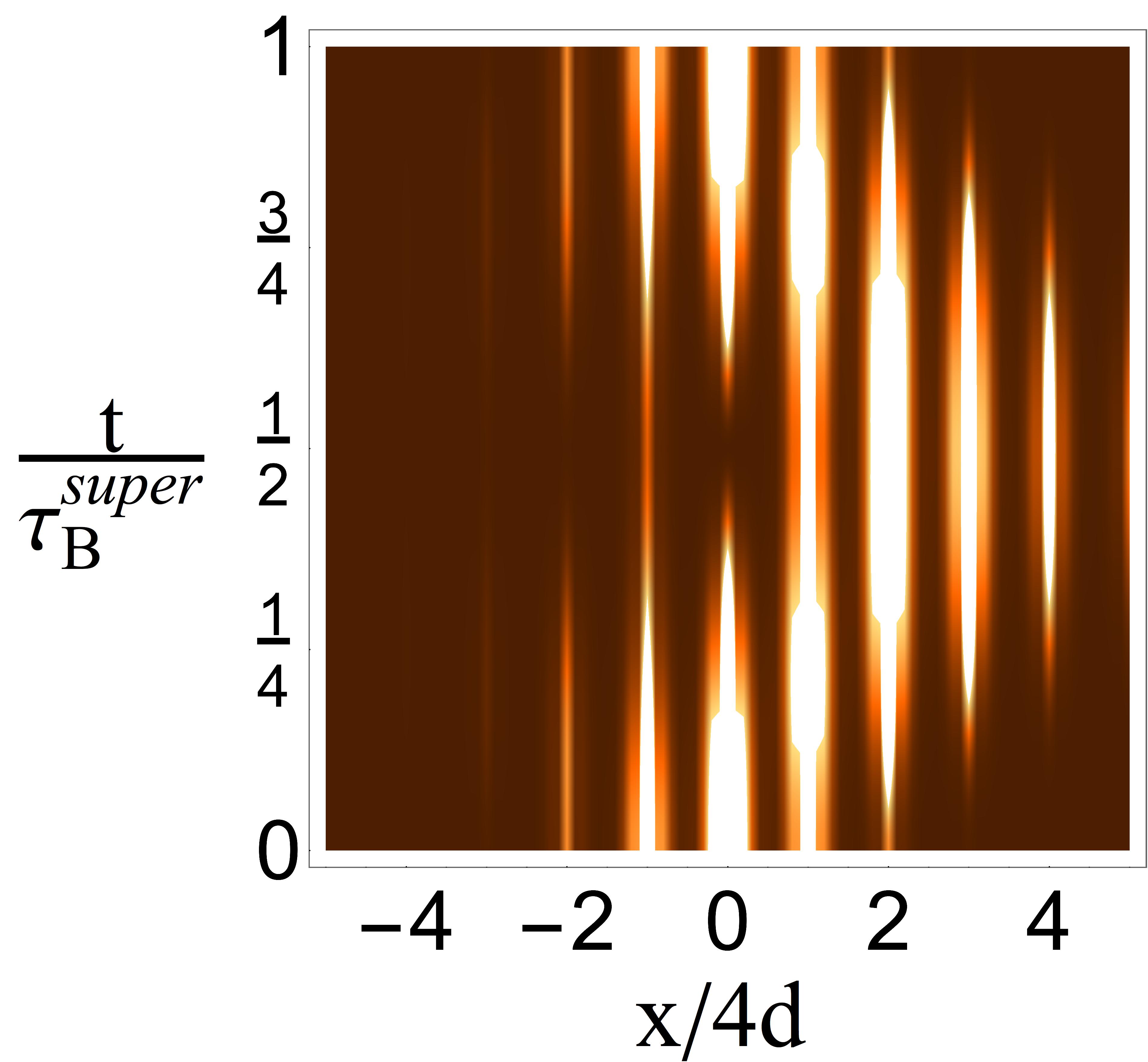}
 \caption{\label{rspacedyn}{\it Colour online.} Time evolution of the wavefunction $|\psi(x,t)|^2$ for one Bloch period when $A_1=3E_R$, $A_2=2E_R$ and $\phi=0$ with $F=1\times10^{-4}E_R/4d$ (left) and $F=2\times10^{-4}E_R/4d$ (right). The initial quasimomentum distribution, localised in the lowest energy band, is described by a Gaussian function centred on $k=0$ with standard deviation $\sigma=2k_r/5$.}
 \end{figure} 
\section{\label{superbo}Bloch oscillations in the superlattice}


 We consider first the case where the force is small enough so that tunnelling to higher energy bands can be neglected.
Figure \ref{rspacedyn} is a simulation of the atomic wave-function, with the same initial spatial distribution shown in Fig.\ref{iwf} in the lowest energy band, evolving for one Bloch period of the superlattice, $\tau_B^{super}=2\hbar\pi/Fd_{super}=\hbar\pi/2Fd$. The plots, from left to right, use $F=1\times10^{-4}E_R/4d \approx 0.7\times10^{-4}mg$ and $F=2\times10^{-4}E_R/4d \approx 1.4\times10^{-4}mg$ respectively. These values are weak enough to ensure no transitions to higher bands occur. From the figure it is clear that when we increase the force the amplitude of the oscillations decreases. 
Bloch oscillations constrain how the wave-function becomes displaced: atoms move to the right due to the direction of our force. These results show that for the atomic cloud to perform Bloch oscillations we need a very weak force, much weaker than gravity, acting on the system.

As mentioned previously, as the value of the force increases we must consider the tunnelling between energy bands. We can simulate these effects by numerically solving Eq.(\ref{cterm}) with the requirement that $\sum_\alpha |c_\alpha|^2=1$. In the following we employ Landau-Zener type tunnelling between energy bands for quasi-general control at sequential passage of avoided crossings within the Brillouin zone. We define the transition probability $T_{\alpha\beta}$ from band $\alpha$ to band $\beta$ as follows: we prepare the atomic cloud such that $|c_\alpha[k(t=0)]|^2=1$ while the others are zero and $T_{\alpha\beta}= |c_\beta[k(t_{max})]|^2$ where $t_{max}$ is the instant of time where $|c_\beta[k(t)]|^2$ is maximum.

\begin{figure}[t]
 \centering
\hspace{-15pt}\includegraphics[scale=.34]{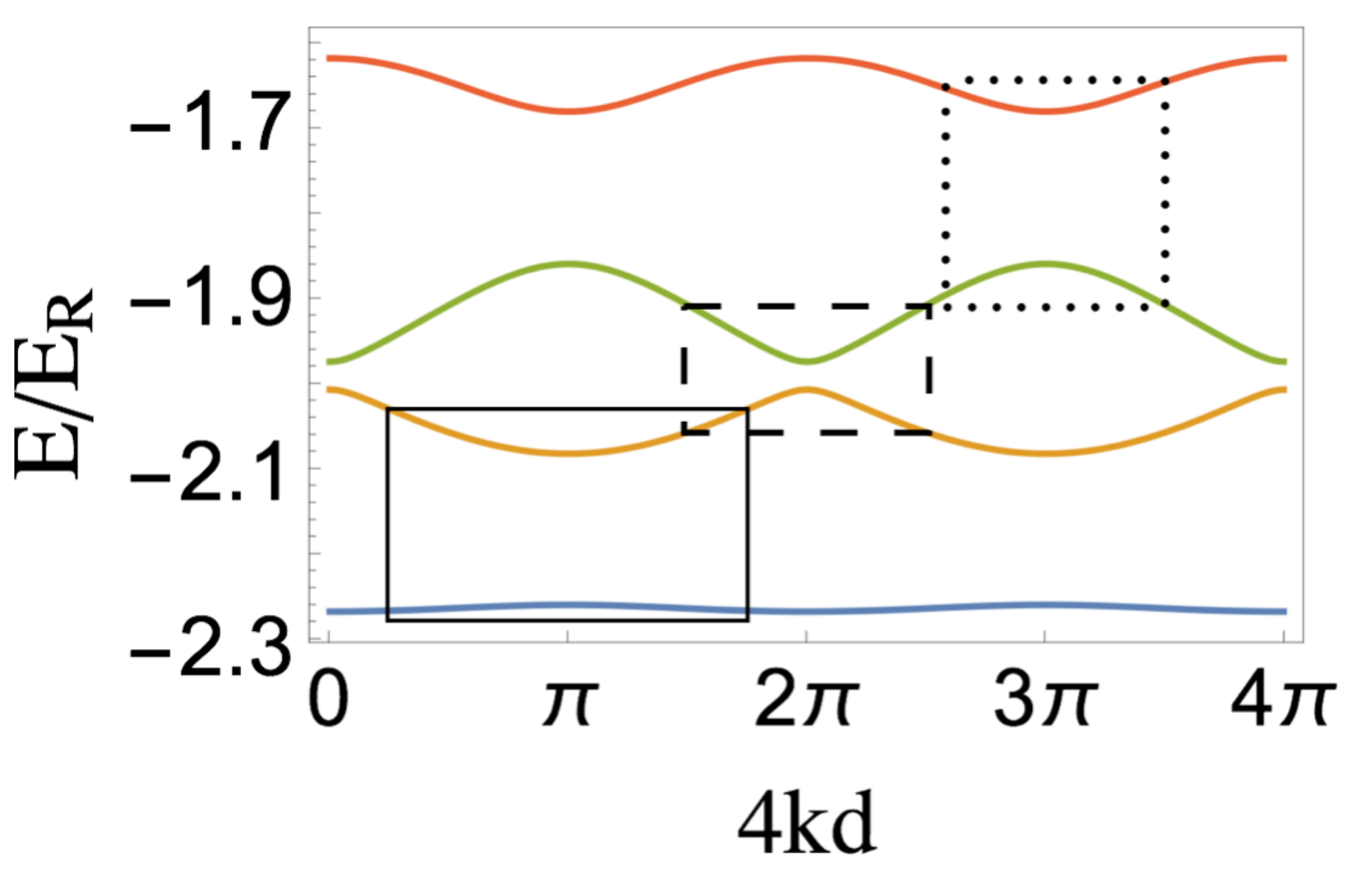}\\
 \includegraphics[scale=0.2]{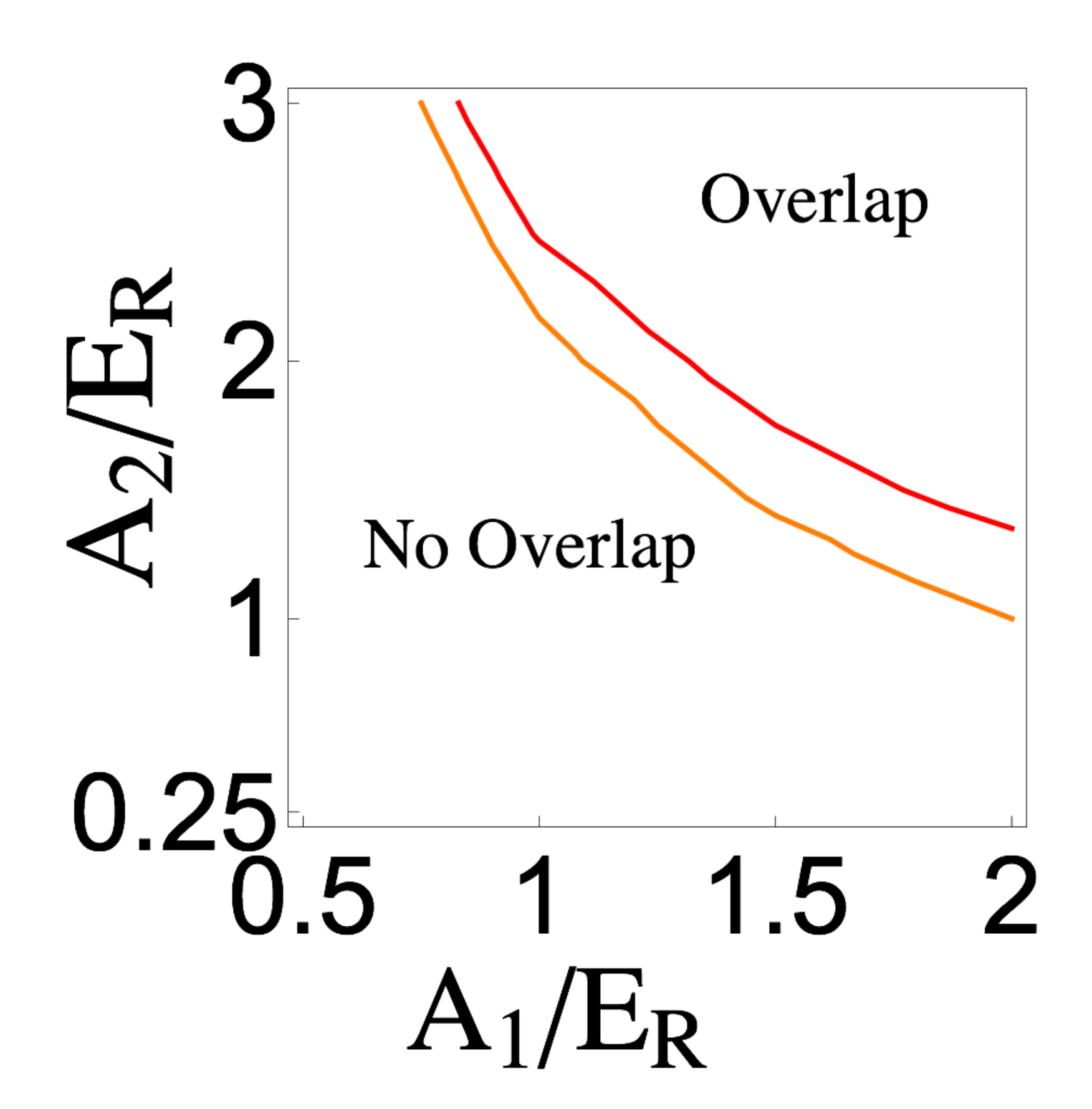}
  \includegraphics[scale=.2]{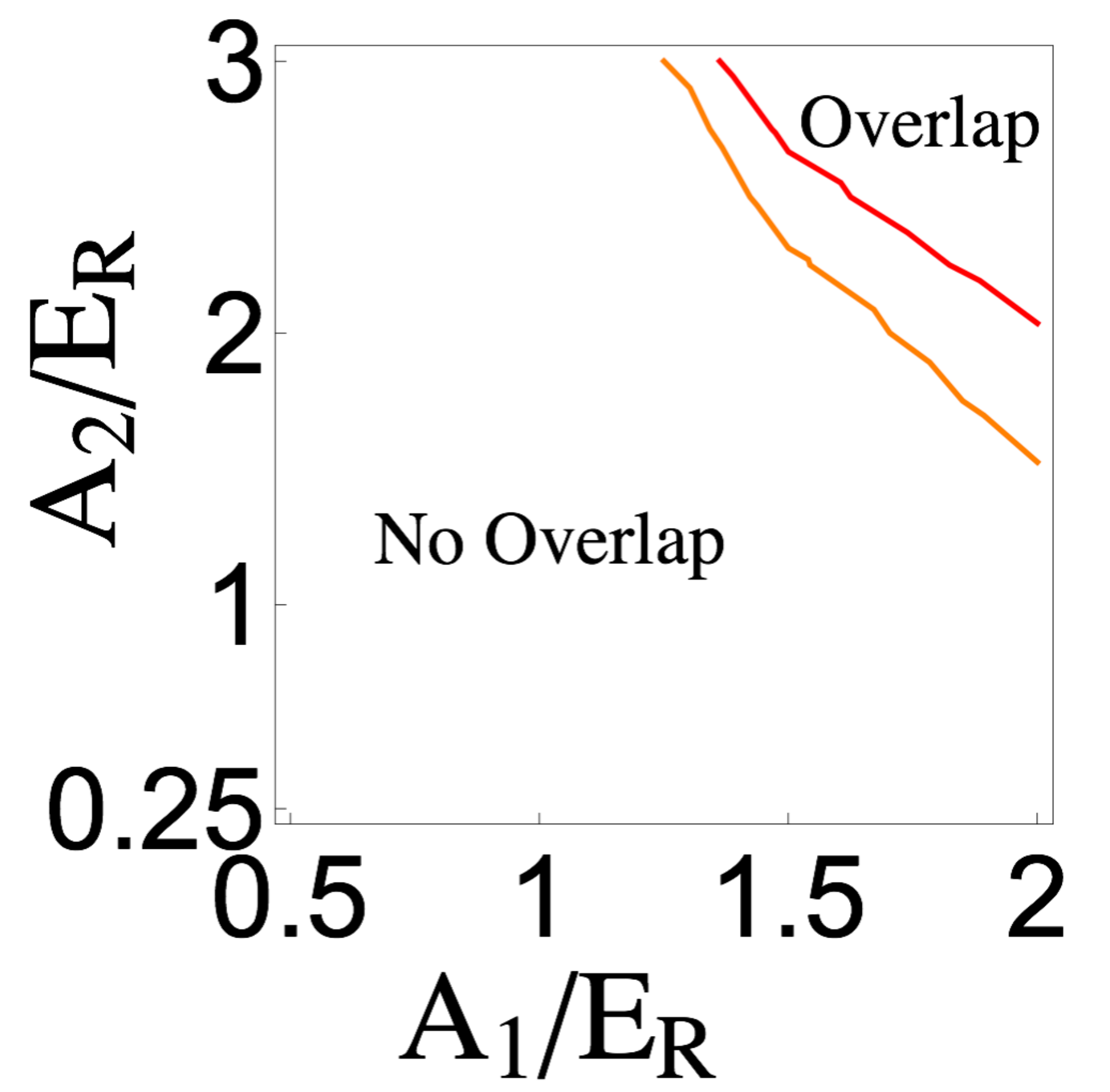}
 \caption{\label{widthsfig}{\it Colour online.}  Shown on top is an example of transition widths: the lowest four energy bands in the superlattice for $A_1=A_2=2E_R$ and $\phi=0$. The widths of the transitions (where interband transitions can occur) are depicted by the black boxes. On the bottom left is a diagram showing the range of parameters of $A_1$ and $A_2$ for which the transitions $1\to 2$ and $2\to 3$ do or do not overlap for $\phi=0$ (red) and $\phi =\pi/8$ (orange). The bottom right is a similar diagram for the transitions $2\to 3$ and $3\to 4$. Here $F=0.05E_R/4d$.}
\end{figure}

Ideally these transitions would be sharply defined, enabling general control as a sequence of two-level beamsplitters \cite{RZt}. However, in reality the transitions have a finite width $2\varepsilon$ \cite{widths} such that they occur in a region $[k_c-\varepsilon,k_c+\varepsilon]$, where $k_c$ is the quasimomentum at the minimum band gap. We define the half-width $\varepsilon$ such that, when the particle's quasimomentum approaches $k_c-\varepsilon$, at least $0.5\%$ of the population of one band transitions to the other. This finite width means that the avoided crossings are not always independent of each other; Fig. \ref{widthsfig} shows the values of $A_1$ and $A_2$ where the avoided crossings overlap. While we wish to avoid or minimise this type of overlap it is possible to utilise it to implement fast reliable transfers between energy levels \cite{tichy, poggi}.

We will consider separately two scenarios for the dynamics in the superlattice: with very low and very high potential depths. For very low potential depths the width of the avoided crossings can, to an extent, be isolated from each other (see the top panel in Fig.~\ref{widthsfig}). For high potential depths, creating flat bands, possible when $A_1,A_2\geq5E_R$, the concept of transition widths becomes irrelevant and population exchange resembles Rabi oscillations instead of Landau-Zener transitions. Although mathematically our treatment is the same, the physics of the two scenarios is significantly different and strongly affects the time scale of the evolution: this is set by the Rabi frequency in the deep lattice regime and by the Bloch period for a shallow lattice. 
To simplify calculations we consider the initial quasimomentum distribution of the particles to be modelled by a Dirac $\delta$ function centered at $k=0$. This is a good approximation for narrow Gaussian distributions centered at $k=0$ with a standard deviation $2k_r/5$ or smaller.

\subsection{Population transfers with low potential depths}\label{lowpot}

 Our aim is to achieve a degree of control over the superlattice without the procedures becoming experimentally difficult, i.e., minimising the changes made to the superlattice. To realize general control we need to be able to design inter-band beam splitters with transition probabilities spanning the interval from $0$ to $1$ for each avoided crossing. Here we present one case where this can be achieved.
 
The parameters we consider are $F=0.05E_R/4d$, $A_1=2E_R$, and $\phi=\pi/8$. We are presenting this case as an example; changing the superlattice parameters, or the force, provides a plethora of accessible transition probabilities. With these experimental settings we numerically measure population transfer across each avoided crossing in the band structure with the exception of the fourth and fifth energy bands. As the distance between these bands is very large, these probabilities quickly fall to zero for $A_1>A_2>1E_R$. 

\begin{figure}[t]
 \centering

  \includegraphics[width=\columnwidth]{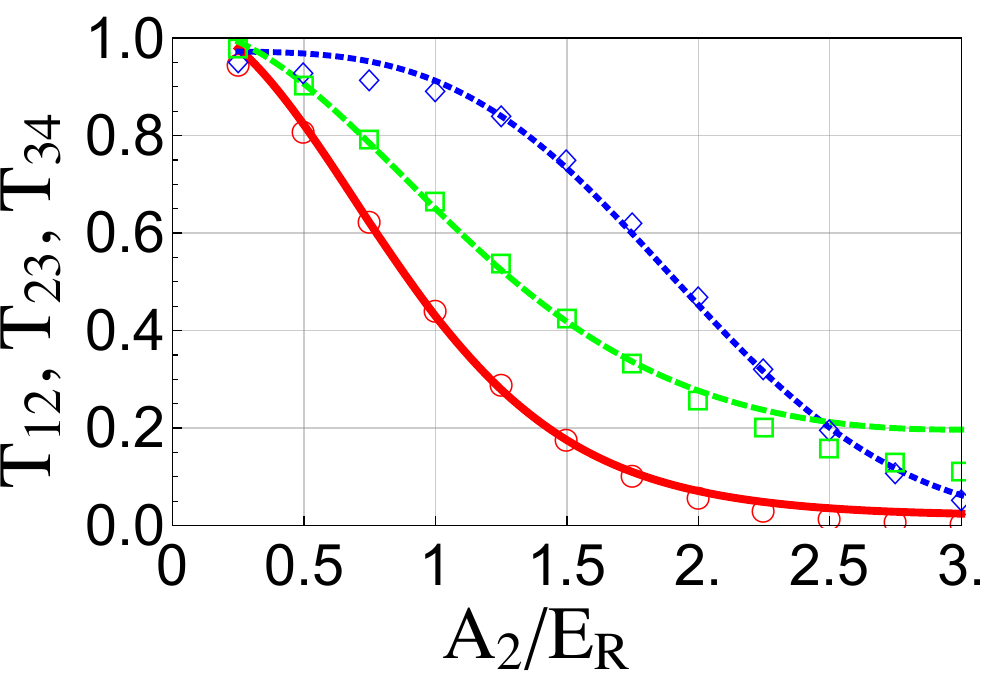}
 \caption{\label{transitions}{\it Colour online.} Example of population transition probabilities at each avoided crossing when $A_1=2E_R$ and $\phi=\pi/8$. Shown are $T_{12}$ (red solid curve with circles) and $T_{34}$ (green dashed curve with squares) measured at $k=\pi/4d$ in the Brillouin zone, and $T_{23}$ (blue dotted curve with diamonds) measured at $k=0$. The curves represent the theoretical predictions for these transitions from the Landau-Zener formula; the markers represent our simulations.}
\end{figure}

Figure \ref{transitions} presents transition probabilities at each avoided crossing in the superlattice band structure: the curves represent probabilities from the analytical Landau-Zener formula and the markers represent our simulations. Here $T_{12}$ (red, solid curve with circles), $T_{23}$ (blue, dotted curve with diamonds) and $T_{34}$ (green, dashed curve with squares) are plotted against $A_2$ ranging from $0.25E_R$ to $3E_R$. For each data point the atomic cloud was prepared in the lower of the two bands a distance $\pi/4d$ from the minimum band gap. From this point the system evolves for a full Bloch period, ensuring that the full transition width is covered. The transition probabilities $T_{12}$ and $T_{34}$ are measured at the same point in the band structure ($k=\pi/4d$) and so their curves are similar. As the minimum band gap between bands 3 and 4 is, on average, smaller than between 1 and 2, the probabilities $T_{12}$ are consistently lower than $T_{34}$ as the potential depths increase. For both curves changing $\phi=\pi/8\to 0$ would decrease the transition probabilities for higher values of $A_2$, as the energy difference between the second and third bands decreases [see Fig.2(b) and 2(f)]. The blue curve ($T_{23}$) is calculated instead at a different location in the band structure (k=0) and, when $\phi=\pi/8$, bands 2 and 3 begin to separate as the potential depth increases. Changing $\phi=\pi/8\rightarrow0$ would significantly increase these probabilities. We have included band structures for the extreme values of $A_2$ shown in Fig. \ref{transitions} to provide some intuition on the avoided crossings at these values in Fig. 2 of the Supplemental Material \cite{supp}.

\begin{table}[b]
\caption{\label{comparetable}Comparing transition probabilities at similar avoided crossings.}
\begin{tabular}{|c|c|c|c|c|}
\hline
&$A_1/E_R$ & $A_2/E_R$ & $\boldsymbol{\phi}$ &$T_{34}$\\ \hline
\multirow{2}{*}{$T_{12}=0.25$}&$1.5$ & $1.69$ & $0$ &$0.606$\\
 & $2.0$ & $1.27$ & $0$ &$0.505$\\\hline
\multirow{3}{*}{$T_{12}=0.5$}&$1.0$ & $1.79$ & $0$ &  $0.868$\\
&$1.5$ & $1.21$ & $\pi/8$  &$0.786$\\
&$2.0$ & $0.92$ & $\pi/8$ &$0.703$\\\hline
\multirow{2}{*}{$T_{12}=0.75$}&$0.5$ & $2.32$ & $\pi/8$ &$0.985$\\
&$1.0$ & $1.15$ & $\pi/8$ &$0.95$\\
\hline
\end{tabular}
\end{table}

As mentioned, when we are performing control processes we need to keep in mind that the $1\rightarrow2$ and $3\rightarrow4$ transitions happen at the same location in the Brillouin zone. In general, these avoided crossings cannot be independently controlled: changing parameters to create a specific transition in one crossing will inevitably affect the other: Table \ref{comparetable} shows a small example, for specific values of $T_{12}$ ($0.25$, $0.5$ and $0.75$ are shown), of which values of $T_{34}$ can be realised. There are certain lattice parameters that allow these transitions to be treated independently, however one cannot achieve a transition probability of $1$ in either of these crossings while the other is $0$.

\subsection{Population transfers with high potential depths}\label{highpot}

Realising flat energy bands in close vicinity to each other can create some interesting phenomena but normally requires engineering the lattice topology in two-dimensional and quasi-one-dimensional lattices~\cite{flatband1, flatband2,flatband3}. In simple lattice structures the energy gap between bands increases with the potential depth so this type of energy band-control is not possible. Normally very exotic lattice geometries are considered to overcome this, here we demonstrate that this behaviour can be realized in our experimentally feasible approach and we explore the potential for quantum state preparation and transfer in this geometry. 

\begin{figure}[t]
 \centering
 \hspace{-15pt} \includegraphics[width=\columnwidth]{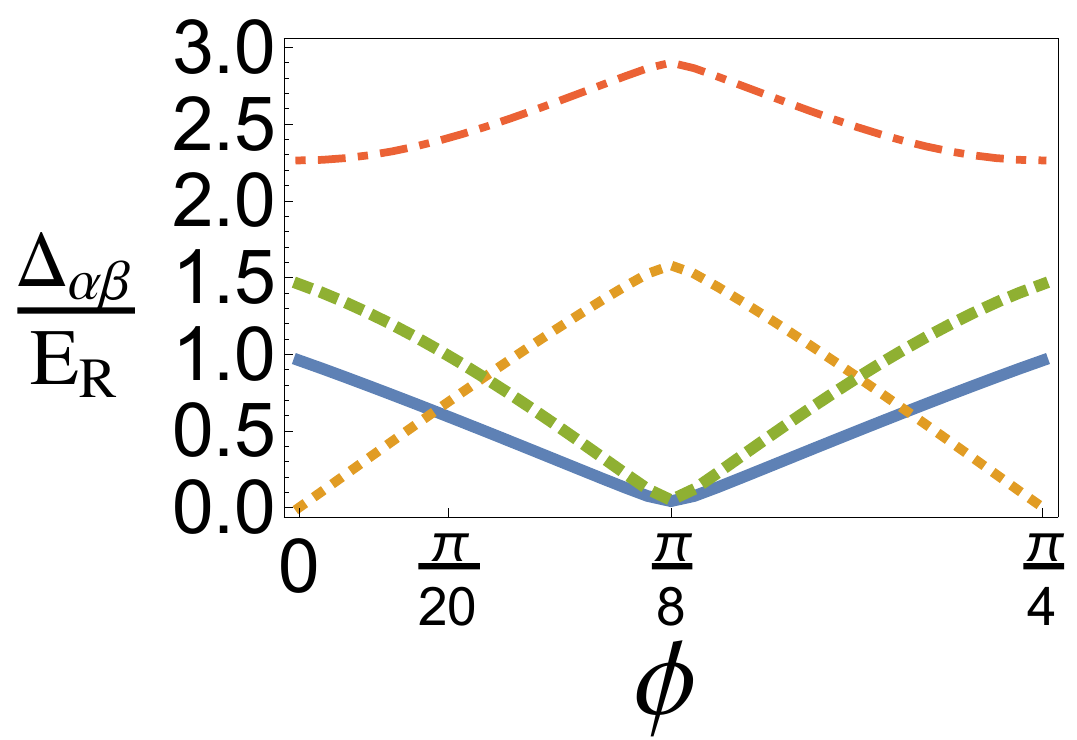}
 \caption{\label{superbsevo} {\it Colour online.} 
Gap between each band ($\Delta_{\alpha\beta}$) when $A_1=8E_R$ and $A_2=5E_R$, corresponding to a band structure with very flat bands, with changing $\phi$. The lines shown are $\Delta_{12}$ (blue solid line), $\Delta_{23}$ (orange dotted line), $\Delta_{34}$ (green dashed line) and $\Delta_{45}$ (red dot-dashed line). }
\end{figure}

By increasing the values of $A_1$ and $A_2$ we can create nearly flat bands, such that the band gap varies only $0.5\%$ from the mean value across the Brillouin zone. To better understand the setup in this regime, and a more intuitive interpretation of the flat band dynamics, the band structures for this scenario can be seen in the Supplemental Material. In comparison to a simple lattice in Fig.~\ref{bspotcomp}, where only the lowest two bands are approximately flat for $V_0=20E_R$, the lowest four energy bands of the superlattice are already flat at $A_1,A_2\geq5E_R$. Figure \ref{superbsevo} shows the superlattice energy band distance $\Delta_{\alpha\beta}$ between bands $\alpha$ and $\beta$ against $\phi$, proving that pairs of energy bands in the superlattice can be made almost degenerate. As mentioned previously, $\phi$ controls the gap between the second and third bands and this plot shows that we can strongly couple energy bands only by changing $\phi$.


\begin{figure}[t]
 \centering
\hspace{-15pt}\includegraphics[width=\columnwidth]{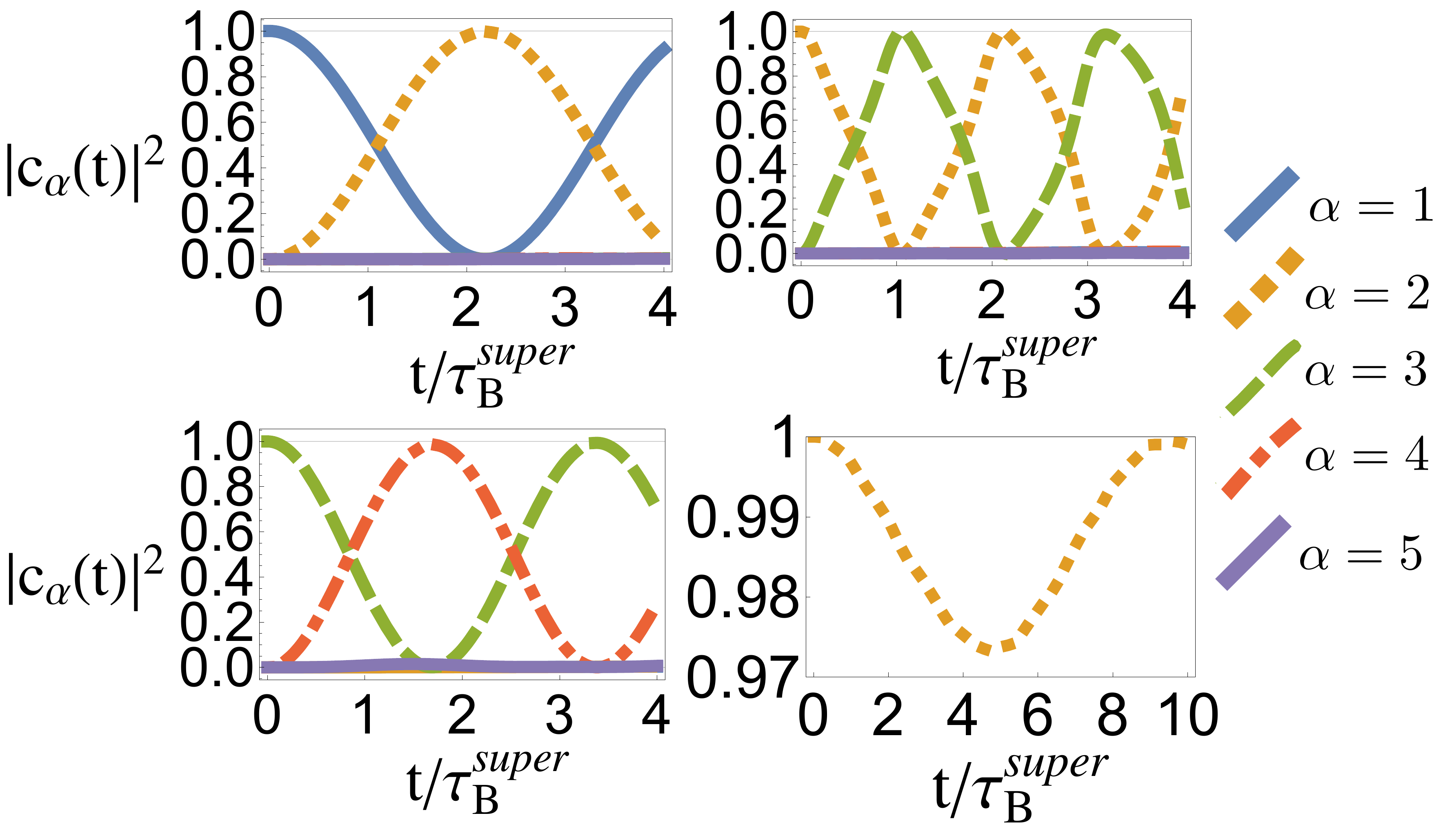}

 \caption{\label{deeptransitions} {\it Colour online.} Tunnelling between energy bands for a very deep potential. Each image has $A_1=8E_R$ and $A_2=5E_R$. The top left shows the population exchange between the first and second energy bands for $\phi=\pi/8$. The top right shows the population transfer between the second and third bands when $\phi=0$. The bottom left shows the transfer between the third and fourth bands with $\phi=\pi/8$. The bottom right shows particles loaded into the second energy band and no two bands are coupled iwth $\phi=\pi/20$. Even after ten Bloch periods we retain $99.89\%$ in the second energy band. Small population transfer to the lowest energy band does occur as the bands are not mathematically flat.}
\end{figure}

Being able to `couple' energy bands in the superlattice band structure means we can also transfer population between bands. In order to encourage this exchange we use a force equivalent to gravity ($\approx 1.42E_R/4d$) to increase the velocity of the particles. For these tests we chose $A_1=8E_R>A_2=5E_R$ to ensure isolation of the lowest four energy levels. The band dynamics in this regime will be considered for three values of the relative phase corresponding to a strong coupling between the second and third energy bands ($\phi=0$), a separation of the energy bands ($\phi=\pi/20$), and strong coupling between the first and second, and between thet third and fourth energy bands ($\phi=\pi/8$). By preparing the particles in the lower band of each pair we let the system evolve for four Bloch periods of the superlattice, aiming to show how population transfers can still occur for very flat bands. Figure \ref{deeptransitions} shows these results. 

The population transfer between the lowest two energy bands (top left) take over two Bloch periods to complete. Between the second and third bands (top right) transitions are much faster as the bands become almost degenerate in energy. The transfer between the third and fourth energy band (bottom left) is slightly faster than in $1\to2$, as the energy gap in the former is consistently lower than the latter. The bottom right panel is an example of how we can suppress population transfers by separating the bands. With the particles initially in the second energy band with $\phi=\pi/20$ even after ten Bloch periods, $99.89\%$ of the population has remained in the second band. There is a small population transfer into the lowest energy level, as the bands are not perfectly flat, although this oscillates. Using a very deep potential, the relative phase can be used to almost totally suppress population transfers over long periods of time. While the bands are very flat there are no longer avoided crossings for the particles to encounter and instead transitions can occur across the entire Brillouin zone. In this sense, these oscillations are very similar to Rabi oscillations.

\section{\label{manipulate}Controlling the energy band populations }
The analysis of the previous section allows us to introduce the main results of the paper. Knowledge of how a BEC divides across avoided level crossings, as well as how the avoided crossings overlap and interact with each other, gives us the ability to manipulate, in an almost general way, the dynamics of an atomic cloud in the superlattice. While we do not consider interference effects here, repeated Landau-Zener transitions leading to interference are reviewed in \cite{Shevchenko}, with a special case of superlattice Landau-Zener tunnelling explored in \cite{breidnew} and experimental realisations in \cite{klingnew, dreisownew, referee3stuckel}.

To perform this control effectively we will change the superlattice parameters {\it in situ} in a step-wise manner. As the Bloch period of the superlattice is of the order $\tau_B^{super}\sim10^{-4}$s when the force present is equal to gravity, changes to the optical lattice structure can be achieved using acousto-optic modulators acting on a timescale of microseconds ($\sim 10^{-6}$s). Consequently we are able to assume that changes to the superlattice parameters occur instantaneously in our simulations. This stepwise method of varying the parameters allows us to simulate quantum state manipulation in the superlattice. This strategy is well suited for the numerical resolution of Eq.~\eqref{cterm}: we assume that the state is frozen while changing the lattice parameters.  
%
The data associated with Fig.~\ref{transitions} allows us to utilise the band structure of the superlattice as a series of concatenated beam splitters, picking our sets of parameters carefully to achieve a state distributed among different bands or localised in one band. With the multitude of control parameters we have a wide range of options for state engineering, and 
we present a few examples below of the potential uses of these concatenated beam splitters and making careful choices of parameter changing. 

\subsection{Moving the condensate from the first to the fourth energy band}
As a first example, we aim at transferring all the population of the first band into the fourth band. This is achieved by sequentially transferring the atomic sample from the first to the second band, then to the third and finally to the fourth energy band. Within a simple lattice structure tunnelling into the higher energy bands is easily done, although the packet will continue to travel up the energy bands until it is no longer trapped inside the lattice. In contrast, the structure of the superlattice allows us to isolate the particles in the fourth band and prevent transitions into higher bands. The large energy gap between the fourth and fifth energy band, as shown in Fig.\ref{bspotcomp}, prevents exchanges at the avoided crossings already for $A_1,A_2>1E_R$.  

\begin{figure}[t]
\centering
\hspace{-83pt} \includegraphics[scale=.7]{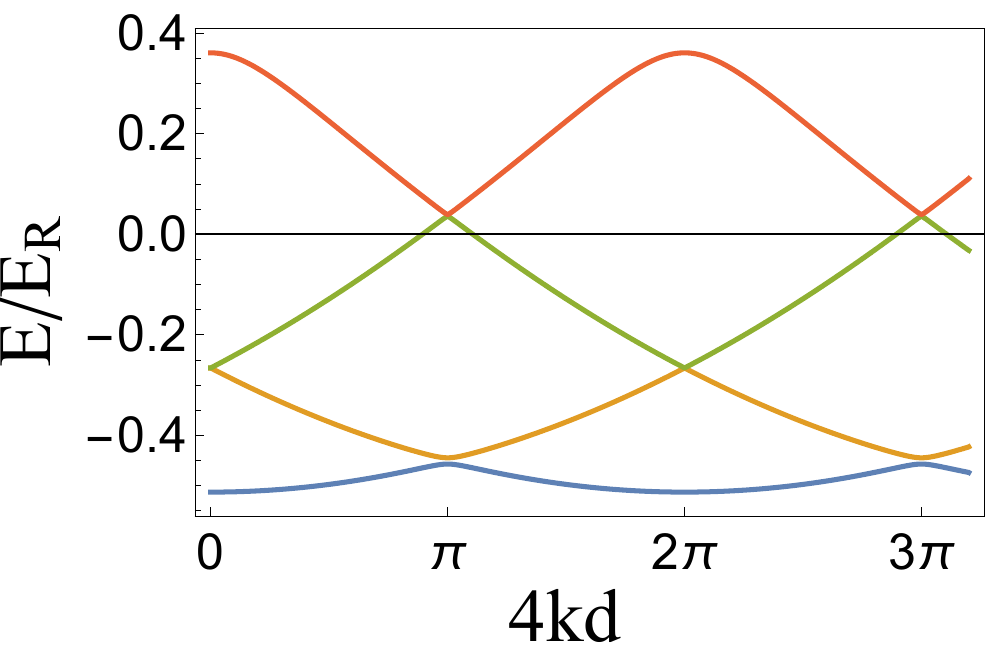}\\
\hspace{-25pt} \includegraphics[scale=0.25]{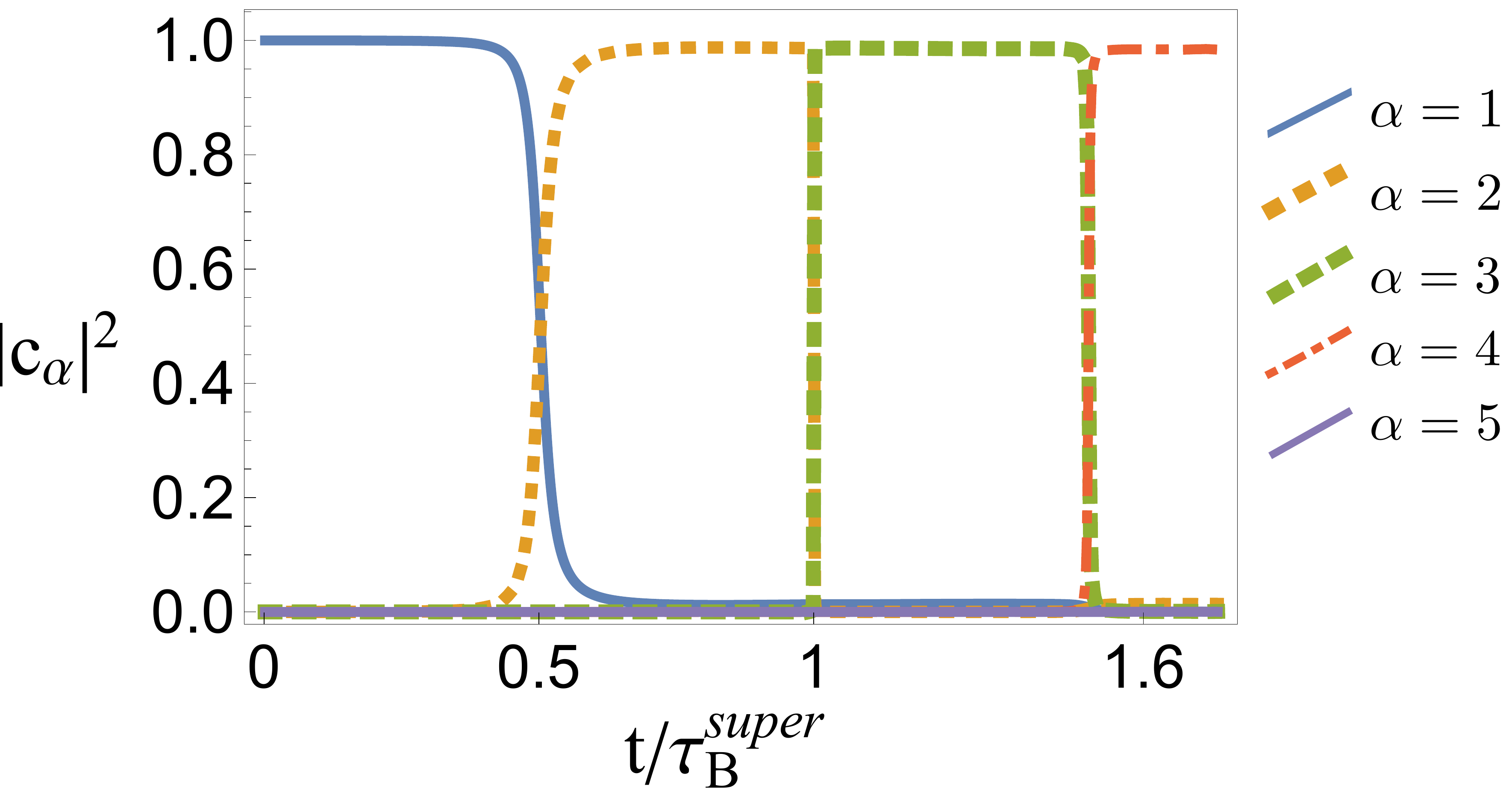}
 \label{paper14}
 \vspace{-10pt}
 \caption{\label{1to4}{\it Colour online.} {\it Top:} The relevant band structure for the simulation in Sect. IV A lasting $1.6 \times \tau_B^{super}$ when $A_1=0.5=2A_2$ and $\phi=\pi/8$. The BEC starts at $k=0$ in the lowest energy band. {\it Bottom:} The band population probabilities plotted against time. The wave packet starts in the first energy band (blue, solid) before making complete transitions into the second band (yellow, dotted), the third (green, dashed) and finally in the fourth energy band (red, dot dashed). The evolution has been extended beyond the final time of $1.6\tau_B^{super}$ to provide clarity on the line styles.}
\end{figure}

We prepare the condensate in the lowest energy band with its quasimomentum distribution modelled by a Dirac $\delta$ function on $k=0$. The force acting on the particles is $0.05E_R/4d\approx 0.035$mg. As can be seen from the figures in Sec. \ref{lowpot}, in order to achieve maximum population exchange at avoided crossings the superlattice must be prepared with very low potential depths, $A_1=0.5E_R$, $A_2=0.25E_R$, and $\phi=\pi/8$. We aim to stop the simulation (i.e. turn off the external force) when the population probability of the fourth energy band is maximum. 
The time this happens at is $t_1=1.6 \tau_B^{super}$ with\[|\mathbf{c}(t_1)|^2=\{0, 0.004, 0, 0.996, 0\}.\] 
Figure \ref{1to4} shows the evolution of the population bands during this process in the bottom panel and the corresponding band structure in the top panel. The energy band gaps at each avoided crossing, perhaps not visible in the plot, are of the order $10^{-2}E/E_R$, $10^{-4}E/E_R$ and $10^{-3}E/E_R$ for the first, second and third avoided crossings encountered by the wave packet respectively. The evolution has been extended beyond $t_1=1.6\tau^{super}_B$ to provide clarity on the line styles.

\subsection{Creating a balanced superposition in the second and third energy band}
Here we show an example of splitting a wave packet initially prepared in the lowest energy band evenly between the second and third energy bands. While it is clear that numerical optimisation would achieve a high fidelity process, we will first develop an intuitive approach. Even if this is perhaps less powerful, it nevertheless provides a deeper insight into the role of each transition in the overall process. Of course, for any practical application numerical optimisation can yield solutions to the required accuracy.

To create this superposition we first need a $100\%$ transfer into the second energy band and a further $50\%$ transfer into the third. To achieve $T_{12}\simeq 1$ the superlattice parameters can be set to $A_1=0.5E_R$, $A_2=0.25E_R$ and $\phi=\pi/8$. A $50\%$ transfer between bands 2 and 3 is slightly more difficult, although we can use the results in Fig.\ref{transitions}, as well as data not presented here, to choose our parameters carefully. We have two options when $\phi=\pi/8$: $A_1=1.5E_R$ and $A_2=2.72E_R$
or $A_1=2E_R$ and $A_2=1.95E_R$.

\begin{figure}[h]
\centering
\vspace{5pt}
\hspace{-75pt} \includegraphics[scale=.74]{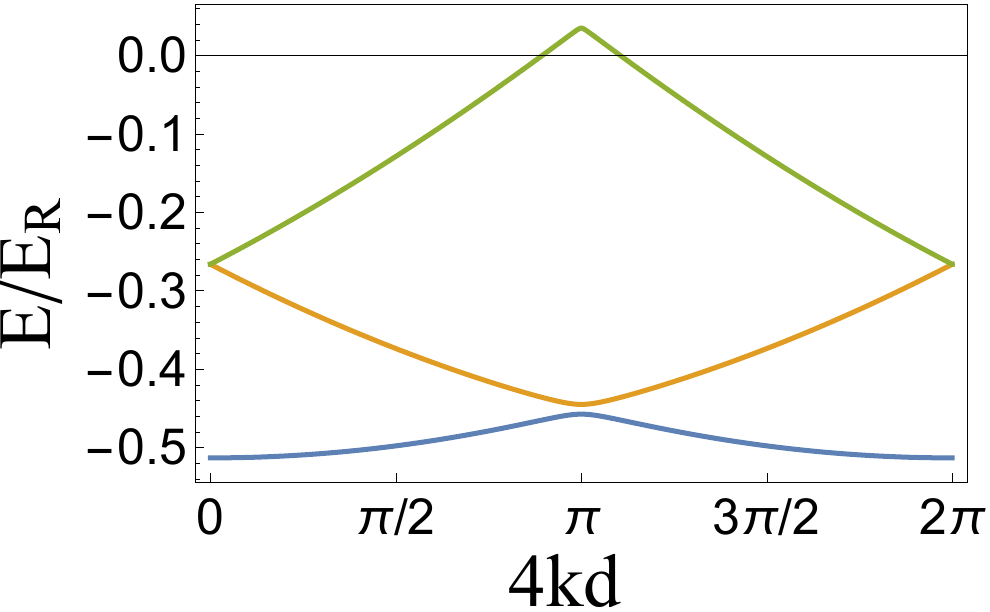}\\
 \hspace{-25pt}\includegraphics[scale=.26]{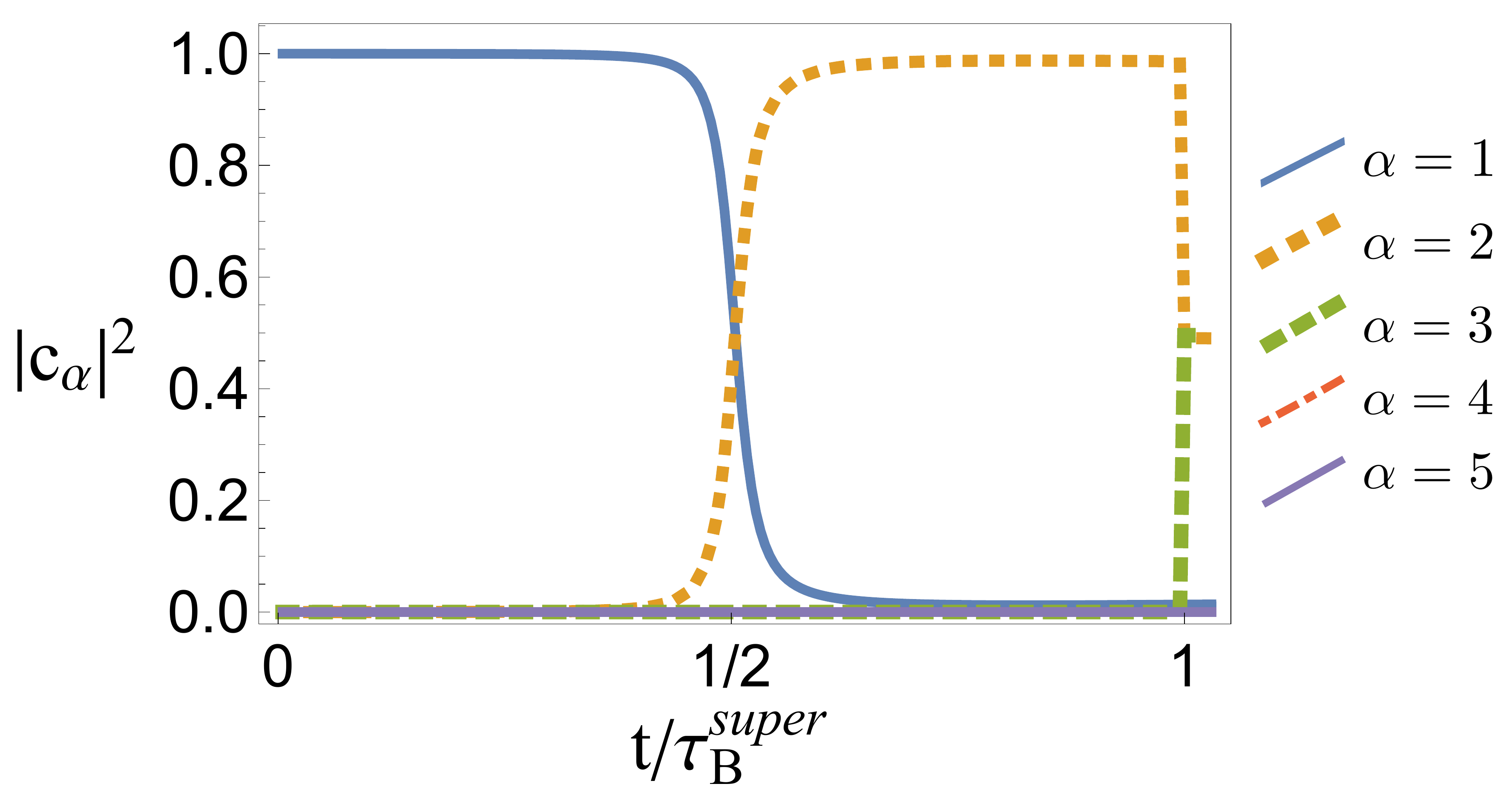}
 \caption{\label{paper23}{\it Colour online.} Shown on top is the relevant band structure for the simulation in Sec. IV B lasting $\tau_B^{super}$ when $A_1=A_2=0.5E_R$ and $\phi=\pi/8$. The bottom shows the band population probabilities plotted against time. The particles are initially populating the lowest energy band (blue solid line) transferring $100\%$ into the second band (yellow dotted line) and transferring $50\%$ into the third band (green dashed line). This plot moves slightly beyond $\tau_B^{super}$ to illustrate the $50/50$ superposition.}
\end{figure}

Recall that these values for the potential depths create a $50\%$ transfer between the second and third energy bands when $\phi=\pi/8$ and $F=0.035$mg: alternative values for these parameters would yield similar, or even superior, results. We have chosen to restrict ourselves to parameters already presented in the paper. It is unavoidable that when we change the parameters to either one of these options the widths of the transitions $T_{12}$ and $T_{23}$ will change and, from Fig.\ref{widthsfig}, overlap. Both of these parameter sets cause the transitions to overlap and we must make a careful choice on the time to implement the parameter changes. From Fig.\ref{widthsfig} it is clear that the lower value of $A_2$ causes the transitions to overlap less, and it is the more convenient choice. This simply reduces the probability of losing some of the particles to the lower band but it does not eliminate it.
Letting Eq.(\ref{cterm}) evolve with the lowest potential depths to a time $t_1=0.89\tau_B^{super}$, right before the minimum band gap between bands 2 and 3, the populations are:
\[|\mathbf{c}(t_1)|^2=\{0.003, 0.996, 0, 0, 0\}.\]  
Continuing the evolution with $A_1=2E_R$ and $A_2=1.95E_R$ (as this gives narrower transition widths than the other) to a time $t_2=1.26\tau_B^{super}$, after the transition, we obtain 
\[|\mathbf{c}(t_2)|^2=\{0.056, 0.505, 0.435, 0, 0\},\]
where some population has dropped down to the lowest energy band. 
We did not achieve a perfect $50/50$ split between energy bands 2 and 3 but it is possible to improve this result. Via careful monitoring of the populations at each time step, we are theoretically able to choose an optimal time for the parameter change. Instead of changing the parameters before the transition between the second and third energy bands we can pinpoint the time where the population transfer is $50\%$ completed. If the superlattice potential depths are set to $A_1=2A_2=0.5E_R$ the avoided crossing between levels one and two is very sharply defined; it becomes difficult to precisely measure the point where the transition is $50\%$ completed. For this reason, we instead set $A_1=A_2=0.5E_R$ such that the dynamics are slightly slower but transition probability is still very high.  Our accuracy of tracking the transition probability in real-time becomes more refined. We see that the $50\%$ transition point occurs at $\tilde{t}_1=\tau_B^{super}$,
\[|\mathbf{c}(\tilde{t}_1)|^2=\{0.014, 0.491, 0.496, 0, 0\}.\] The top panel in Fig.\ref{paper23} shows the relevant band structure up to $\tilde{t}_1$, one Brillouin zone. The bottom panel shows the population evolution from the first to the third band, where for illustrative purposes we continued the evolution slightly further than $\tilde{t}_1$ to show the 50/50 split.

\subsection{An arbitrary split between the first and third energy band}
Due to the coupling of the avoided crossings at $k=\pi/4d$ we have some restrictions when we want to create an {\it arbitrary} superposition in the superlattice. As an example of the power of these concatenated beam splitters here we aim for a final population distribution of $\{0.25, 0, 0.75, 0, 0\}$. We need $T_{12}=0.75$ and $T_{23}=1$ to complete this. This process is done over two Bloch periods. With the analysis presented in Fig.(\ref{transitions}), we know we can have a value $T_{12}=0.75$ when $A_1=2E_R$, $A_2\simeq0.584E_R$ and $\phi=\pi/8$. A $100\%$ transfer between bands 2 and 3 is achieved by extremely low potential depths and in neither case do the transitions overlap. After the first population exchange, at a time $t_1=0.72 \tau_B^{super}$, the population distribution is
\[|\mathbf{c}(t_1)|^2=\{0.25, 0.75, 0, 0, 0\}.\]  We change the parameters to $A_1=0.5E_R$, $A_2=0.25E_R$ and $\phi=0$ and continue the evolution through the avoided crossing between bands 2 and 3, before stopping the evolution at a time $t_2=1.2\tau_B^{super}$. The final population distribution is 
\[|\mathbf{c}(t_2)|^2=\{0.249, 0.002, 0.749, 0, 0\}.\] This example of using the band structure as a series of beam splitters between energy bands might have improved fidelity when optimised. 

 \begin{figure}[t]
\centering
 \includegraphics[scale=.35]{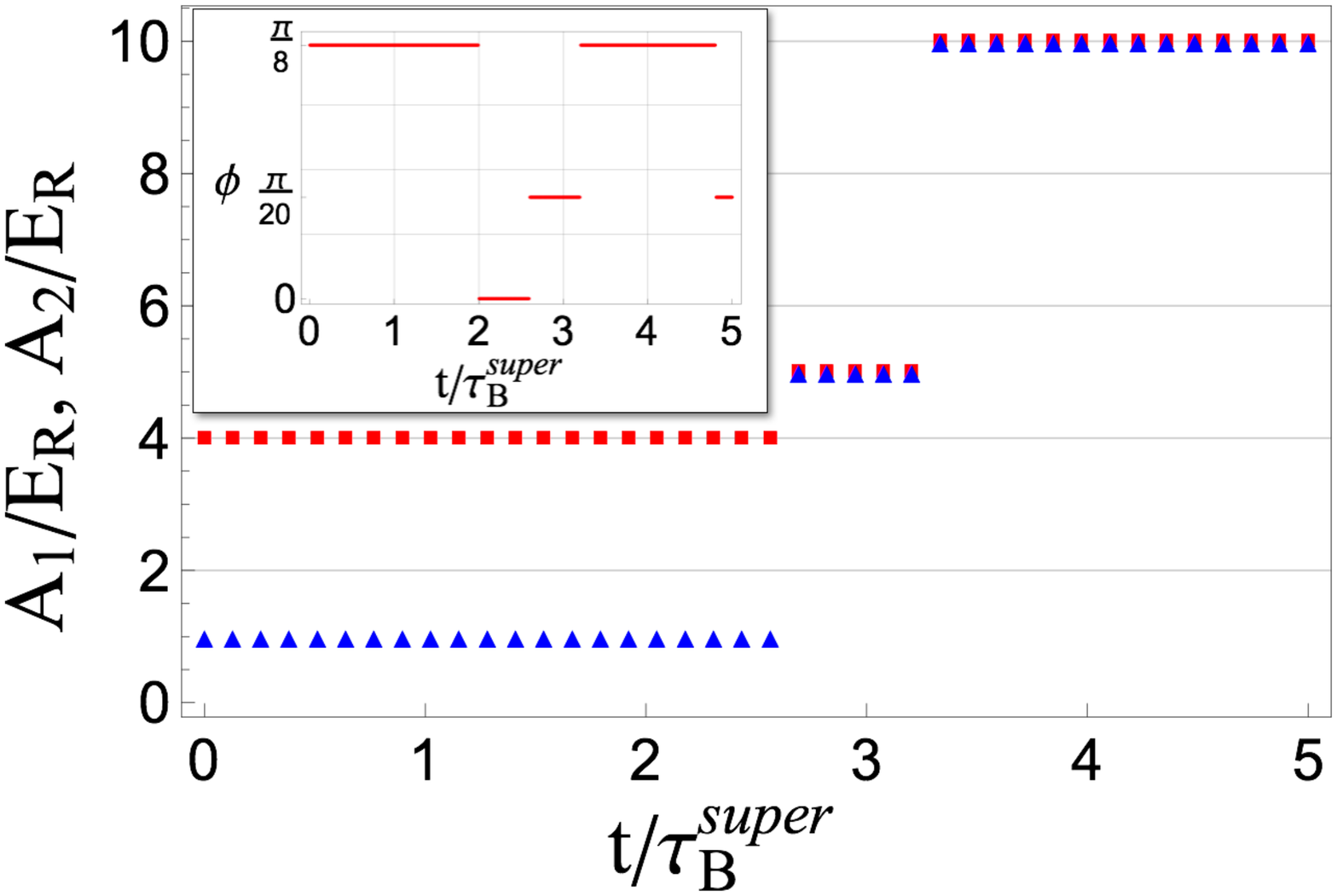}
 \caption{ \label{combo}{\it Colour online.} Values of superlattice parameters we use for the process in Sec. \ref{4d}:$A_1$ (red squares) and $A_2$ (blue triangles) over time. The potential depth changes are performed at $t/\tau_B^{super}=2.6$ and $3.2$. The inset shows the relative phase $\phi$ versus time. The changes are performed at $t/\tau_B^{super}=2$, $2.6$, $3.2$ and $4.8$.}  
\end{figure}
\subsection{From an equal band superposition to a single band}\label{4d}
In our last example of the control over the superlattice, we simulate a wave packet initially distributed across the lowest four energy bands, attempting to recollect it into the lowest energy band, something that is impossible in conventional lattices due to the unavoidable coupling to higher bands.

We use a different value for the force here, $F=0.1E_R/4d$, although this process would be possible for any force of the same order of magnitude. This process may not be possible for larger values of an external force as the internal velocity of the particles can create some  complicated tunnelling dynamics.  The particles are prepared into the lowest four energy bands in a five band approximation such that the initial conditions of Eq.(\ref{cterm}) are
\[|\mathbf{c}(0)|^2=\{0.25,0.25,0.25,0.25,0\},\]
with a Dirac $\delta$ initial quasimomentum distribution centered on $k=0$. This process is conducted over five Bloch periods and we implement four sets of parameter changes. The first step is to empty the fourth energy band of its population into the third. Using the relative phase $\phi$, we can couple the second and third energy bands while isolating the fourth and waiting until the third energy band is empty before changing $\phi$ again and coupling the first and second energy bands. When the second band is empty (and, as a consequence, the population is in the lowest energy level) we isolate the bands from each other and the evolution can stop. The final population distribution is \[|\mathbf{c}(5\tau_B^{super})|^2=\{0.9575,0.0026,0.0104,0.029,0\}.\] We emphasise that other lattice parameters and force values may achieve similar or, by employing optimisation algorithms, better fidelity. 

\begin{figure}[t]
\centering
 \hspace{-30pt}\includegraphics[scale=.43]{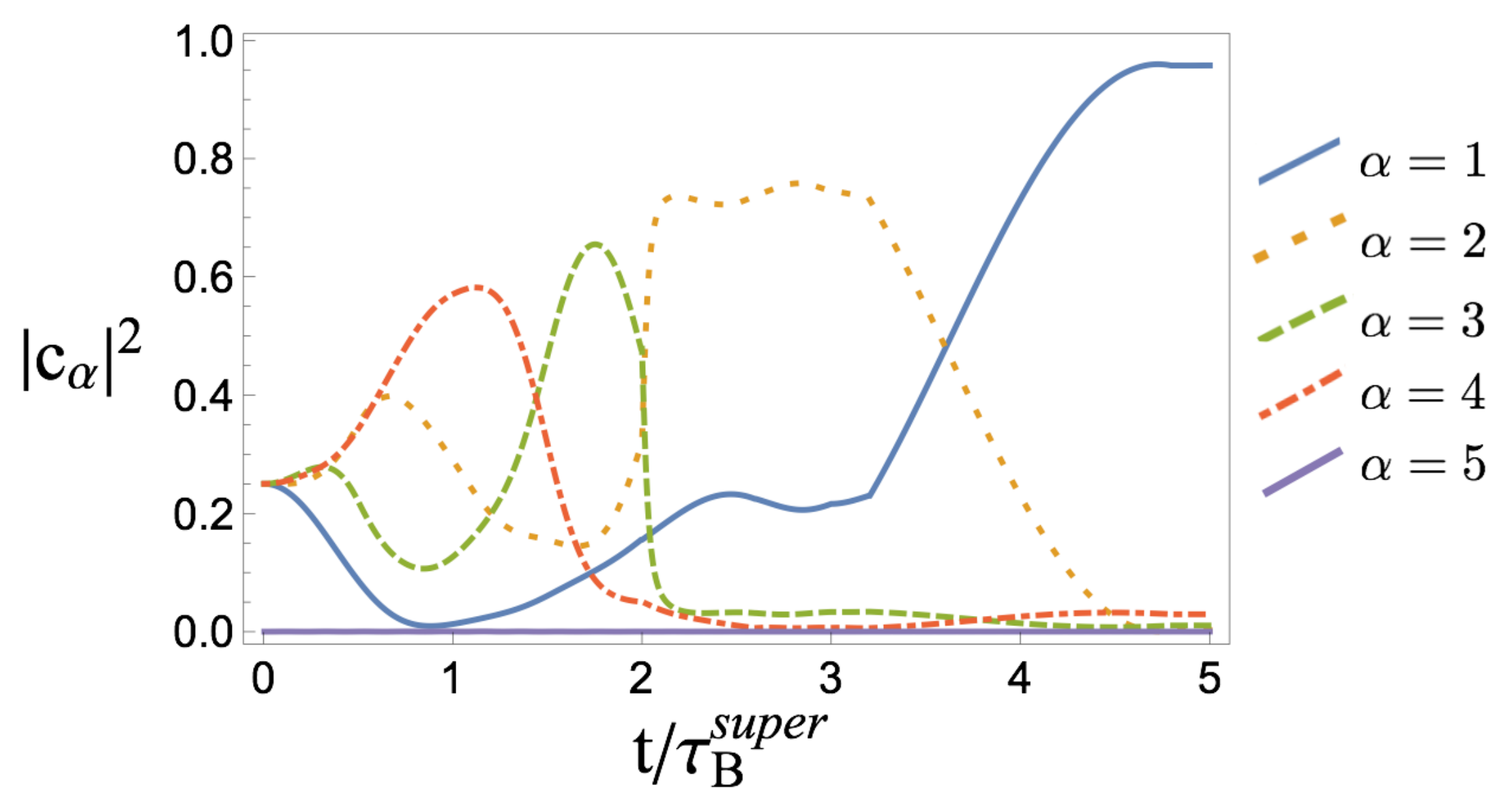}
 \caption{ \label{paperq1}{\it Colour online.} Population probabilities of each energy band plotted against time in the simulation in Sect IV D. The wave packet begins in a state evenly spread across the lowest four energy bands and we dynamically manipulate its movement through the superlattice to regain $96\%$ into the lowest energy band (blue solid line). Time evolution of the second energy band population is shown in the yellow dotted line, the third energy band in the green dashed line, and the fourth in the red dot-dashed line. }
\end{figure} 

Figure \ref{combo} shows the superlattice parameter values we used for this process and at what time they were implemented. The inset details the values of $\phi$ while the main graph shows the potential depth changes. These values for the potential depths are low enough to be easily implemented experimentally. We change $\phi$ more often due to the influence it has over the distance between energy bands 1 and 2. The value $\phi=\pi/20$ is used as a `freezing' value as the bands become almost equally separated.

Figure \ref{paperq1} shows the absolute square value of the solutions to Eq.(\ref{cterm}) plotted against time as the wave packet evolves over five Bloch periods. Beginning in an equal superposition in the lowest four bands the dynamics are manipulated to recollect the wave packet into the lowest energy band.

We have described in Sec. IV various possibilities of band population manipulation in the superlattice. We have not discussed the coherences between each energy band that, through constructive and destructive interference mechanisms, affect the dynamics of the particles. We wish to stress that manipulating band population necessarily involves manipulation of the coherences, and are fully included in our numerical simulations. Using time-of-flight measurements, both of these features (band population and coherence) can be measured, by either suddenly turning off the lattice or band-mapping using appropriate beam splitter transformations \cite{billphillips}.

\section{\label{conc}Conclusions}
In this work we have theoretically explored the manipulation of a non-interacting BEC modelled by a wave packet in a superlattice structure. We employed Bloch oscillations, easily implemented in current experiments, in conjunction with multiple superlattice parameters to manipulate the BEC band populations.
 The ability to create quasi-isolated band multiplets and tunnelling between flat energy bands give clear advantages over a simple lattice. We have showcased a few examples of the available control in this setup with step-wise constant parameters. Higher quality control could be achieved with more general time dependences.  

While we consider the preparation of any arbitrary quantum state in this superlattice to be an open question, utilising exhaustive numerical optimisation a wide range of different quantum states can be created. Superlattice potential structures have already been shown to aid atomic transport \cite{salger} and the use of Bloch oscillations may improve this process. Furthermore, the effect of interatomic interactions in a superlattice was explored in \cite{korschnew} and if, for instance, interactions were controlled using a Feshbach resonance this may provide an additional tool for the manipulation of particles. Their exact role in this problem remains to be analysed.

\acknowledgements \noindent This work was supported by the UK EPSRC, John Tem- pleton Foundation (Grant No. 43467), the EU Collaborative Project TherMiQ (Grant Agreement No. 618074), Professor Caldwell Travel Studentship, the Lundbeck Foundation (Grant No. R139-2012-12633), and the European Research Council (Grant No. 639560)

\bibliographystyle{apsrev4-1}
\bibliography{bib2}
\end{document}